\documentclass[a4paper,11pt]{article}
\pdfoutput=1
\usepackage{jheppub}

\usepackage{mathtools}
\usepackage{physics}
\usepackage{bm}
\usepackage{bbm}
\usepackage{slashed}
\usepackage{cancel}
\usepackage{dcolumn}
\usepackage{float}
\usepackage[title]{appendix}
\usepackage[normalem]{ulem}
\usepackage{comment}
\usepackage{xcolor}

\usepackage{tikz}
\usepackage{tikz-3dplot}
\usetikzlibrary{
  arrows.meta,
  patterns,
  positioning,
  shapes.geometric,
  decorations.pathreplacing
}

\definecolor{violet}{cmyk}{.72,.92,.1,.00}
\definecolor{linkorange}{RGB}{238,154,127}
\definecolor{linkgreen}{RGB}{59,132,35}

\hypersetup{
  colorlinks=true,
  linkcolor=linkorange,
  anchorcolor=black,
  citecolor=linkgreen,
  urlcolor=blue
}

\begin{document}
\count\footins = 1000

\title{Helicity Decay to Baryons: The Cosmological Evolution of $L_\mu-L_\tau$ Gauge Fields and Baryogenesis}

\author[a,b,c]{Kohei Kamada,}
\author[d]{Fumio Uchida,}
\author[a,e,f]{and Di Wu}

\affiliation[a]{School of Fundamental Physics and Mathematical Sciences, Hangzhou Institute for Advanced Study, University of Chinese Academy of Sciences (HIAS-UCAS), 310024 Hangzhou, China}
\affiliation[b]{International Centre for Theoretical Physics Asia-Pacific (ICTP-AP), Hangzhou/Beijing, China}
\affiliation[c]{Research Center for the Early Universe, The University of Tokyo, Bunkyo-ku, Tokyo 113-0033, Japan}
\affiliation[d]{Cosmology, Gravity, and Astroparticle Physics Group, Center for Theoretical Physics of the Universe, Institute for Basic Science (IBS), Daejeon, 34126, Korea}
\affiliation[e]{Institute of Theoretical Physics, Chinese Academy of Sciences, Beijing 100190, China}
\affiliation[f]{University of Chinese Academy of Sciences, Beijing 100049, China}

\emailAdd{kohei.kamada@ucas.ac.cn}
\emailAdd{fumiouchida@ibs.re.kr}
\emailAdd{wudi24@mails.ucas.ac.cn}

\date{\today}

\abstract{
We study baryogenesis from the post-inflationary decay of helical U(1)$_{L_\mu-L_\tau}$ gauge fields generated during axion inflation. Assuming that Schwinger pair production is suppressed during inflation, a large helicity-to-entropy ratio $\eta_H$ can be generated while keeping the initial lepton asymmetry just after inflation negligible. After reheating, the dissipation of magnetic helicity sources a lepton asymmetry through the chiral anomaly, which is  partially converted into a baryon asymmetry by electroweak sphaleron processes. 
We find that there are parameter spaces where the observed baryon asymmetry can be generated in both the turbulent and dissipation-dominated regimes of the U(1)$_{L_\mu-L_\tau}$ magnetic field evolution. The baryon asymmetry generation in the turbulent regime is, however, dominated by the pre-turbulent nearly frozen regime. We further find that the  requirement of successful baryogenesis imposes stronger constraints on the inflationary Hubble parameter and the helicity density generated during axion inflation than the requirement for 
the gauge-field energy not to backreact on the inflaton dynamics.}
\maketitle

\section{Introduction}

The observed baryon asymmetry of the Universe, $\eta_B^{\mathrm{ obs}}\simeq 8.6\times10^{-11}$~\cite{Planck:2018vyg}, is difficult to realize in the history of the universe within the Standard Model of particle physics (SM) ~\cite{Sakharov:1967dj,Dolgov:1991fr} and therefore requires new physics.
Among the proposed mechanisms, the axion inflation provides a remarkable framework for baryogenesis~\cite{Anber:2015yca,Adshead:2016iae,Jimenez:2017cdr,Domcke:2018eki,Domcke:2019mnd}.  With a Chern--Simons coupling between the axion-like inflaton and a U(1) gauge field, one helicity mode of the gauge field is exponentially amplified during inflation~\cite{Anber:2006xt}.  The produced gauge field carries the $\rm U(1)$ Chern--Simons number, whose evolution is related by the chiral anomaly to fermion number production. 
Together with the subsequent post-inflationary evolution, the baryon asymmetry of the Universe can be induced and survive until today~\cite{Giovannini:1997eg,Anber:2015yca,Adshead:2016iae,Fujita:2016igl,Kamada:2016eeb,Kamada:2016cnb,Cado:2016kdp,Jimenez:2017cdr,Domcke:2018eki,Domcke:2019mnd,Domcke:2022kfs}.  

When the gauge field is the SM hypercharge (U(1)$_Y$) field, however, the scenario may suffer from a cancellation between the fermionic asymmetry and the helicity, which suppress the final baryon asymmetry~\cite{Domcke:2018eki,Domcke:2019mnd}.\footnote{
    This problem can be avoided by employing the wash-in mechanism~\cite{Domcke:2020quw,Domcke:2022kfs}.
}\footnote{
    The efficiency of the baryogenesis  from the U(1)$_Y$ helicity decay is also 
    under discussion~\cite{Boyer:2025jno,Hamada:2025cwu,Fukuda:2025nmc}.
}
One may instead consider a U(1) gauge field beyond the SM whose helicity is anomalously linked to $B-L$ in the visible sector.  
In this case, similarly to standard leptogenesis~\cite{Fukugita:1986hr}, 
an induced lepton asymmetry can be partially reprocessed into a baryon asymmetry by the electroweak sphalerons~\cite{Klinkhamer:1984di,Kuzmin:1985mm}; 
see {\it e.g.,} Ref.~\cite{Fukuda:2024pkh}. 
An interesting model in this context is the gauged U(1)$_{L_{\mu}-L_{\tau}}$ one, in which the U(1)$_{L_{\mu}-L_{\tau}}$ helicity can source 
the lepton (flavor) asymmetries through the chiral anomaly.   
In particular, if a large U(1)$_{L_{\mu}-L_{\tau}}$ helicity-to-entropy 
is generated during inflation, it can eventually be converted 
into a large lepton asymmetry after the electroweak symmetry breaking, potentially resolving the tension between the recent ACT results and some inflation models \cite{Hu:2025nep}.
In the present work, we consider 
the same class of post-inflationary initial conditions as of Ref.~\cite{Hu:2025nep}, but focus instead 
on the history of baryogenesis before the electroweak sphaleron freeze-out. 

A distinctive feature of the U(1)$_{L_{\mu}-L_{\tau}}$ scenario is that a large lepton asymmetry can be generated without necessarily overproducing the baryon asymmetry, since the chiral U(1)$_{L_{\mu}-L_{\tau}}$ anomaly violates lepton number but not baryon number. This is in contrast with the previous studies involving the U(1)$_Y$ or the gauged U(1)$_{B-L}$ 
anomaly, where the resulting baryon and lepton asymmetries are tied by order-one chemical-equilibrium coefficients. 
This is because the generated asymmetries before electroweak symmetry breaking is redistributed by Yukawa interactions and sphaleron processes~\cite{Khlebnikov:1988sr,Harvey:1990qw,Nardi:2005hs,Davidson:2008bu,Benso:2026csa}.  
After the sphaleron freeze-out, a pure lepton asymmetry cannot be generated in these scenarios: the hypercharge field transforms into
the electromagnetic field, which does not anomalously violate vector $B$ or $L$, while the decay of a U(1)$_{B-L}$ helicity produces $B-L$ asymmetry rather than 
a pure leptonic (flavor) asymmetry.  

The U(1)$_{L_\mu-L_\tau}$ gauge field studied here is different because its helicity is associated with a flavor-specific lepton asymmetry.  
In the limit of large U(1)$_{L_\mu-L_\tau}$ conductivity, the U(1)$_{L_\mu-L_\tau}$ helicity conserves exactly, and hence no baryon asymmetry is generated if the helicity decay associated with its symmetry breaking occurs after electroweak sphalerons have frozen out. 
In reality, however, the helicity is continuously dissipated and keep decreasing even before the higgsing by the finite U(1)$_{L_\mu-L_\tau}$ conductivity.  
Before the sphaleron freeze-out, part of the helicity decrease can be converted into the observed baryon asymmetry by the electroweak sphalerons. 
The helicity that remains until the sphaleron freeze-out, on the other hand, 
can subsequently generate a large lepton asymmetry without affecting baryon number. 

In the minimal setup for the U(1)$_{L_\mu-L_\tau}$ gauge field production during axion inflation, the accompanying electric field is also generated and can produce the charged fermion pairs through the Schwinger effect.  The induced current backreacts on the gauge field and limits the amount of helicity that survives inflation, thereby restricting the asymmetry obtainable in this scenario~\cite{Domcke:2018eki}.  Reference \cite{Hu:2025nep} proposed a setup to evade this limitation by making the relevant charged fermions heavy during inflation, thereby suppressing the Schwinger pair production.   
This allows a large U(1)$_{L_\mu-L_\tau}$ helicity-to-entropy ratio to be generated while leaving a negligible fermionic asymmetry at the end of inflation. We adopt this class of post-inflationary initial conditions in the present work.

In this work, we examine baryon production from the early-time dissipation of this helicity which was not explored in Ref.~\cite{Hu:2025nep}. The U(1)$_{L_\mu-L_\tau}$ helicity is dissipated before the electroweak sphaleron freeze-out and thereby sources
lepton number through the chiral anomaly. The resulting lepton asymmetry is then partially converted into baryon number by the electroweak sphalerons.  We determine the parameter region in which this process reproduces the observed baryon asymmetry.

To this end, we follow the post-reheating evolution of the U(1)$_{L_\mu-L_\tau}$ magnetic field in the radiation-dominated plasma.  The finite plasma conductivity drives helicity dissipation~\cite{Giovannini:1997gp,Giovannini:1997eg}, while the chiral magnetic effect (CME) provides a washout channel by converting the chiral asymmetry back into magnetic helicity~\cite{Kamada:2016eeb}.  For the magnetic-field evolution, we use the MHD scaling classification of Ref.~\cite{Uchida:2024ude}.  We then solve the evolution equations for the fermion number density-to-entropy ratios, including the SM Yukawa interactions, the strong and electroweak sphalerons, the helicity decay, and the CME feedback.  For fixed coupling $g_{L_\mu-L_\tau}=3\times10^{-4}$, we determine the contour in the $(\eta_H,H_{\rm inf})$ plane along which $\eta_B=\eta_B^{\rm obs}$ is realized. We find that this contour is mainly controlled by either diffusion-dominated evolution or a frozen MHD stage prior to the turbulence.  In much of the relevant parameter space, the turbulent inverse-cascade regimes are not reached before the electroweak sphaleron freeze-out. 

The remainder of this paper is organized as follows. We begin by reviewing the production of helical U(1)$_{L_\mu-L_\tau}$ gauge fields during axion inflation  in Sec.~\ref{Sec:magnetogenesis}. This specifies the post-inflationary initial conditions. In Sec.~\ref{Sec:Baryogenesis}, we describe the subsequent MHD evolution of the magnetic field and couple helicity dissipation to the evolution of the chemical potentials of the SM particles. We then solve the resulting Boltzmann equations and determine the parameter region that reproduces the observed baryon asymmetry. We summarize our results and discuss future directions in Sec.~\ref{Sec:Discussion}. 

\section{U(1)$_{L_\mu-L_\tau}$ magnetogenesis in the slow-roll axion inflation} \label{Sec:magnetogenesis}

In this section, we briefly review the model that we consider. In particular, we explain the amount of a helical magnetic field generated through the Chern--Simons coupling in the slow-roll regime. We also quantify the generated magnetic helicity density as the initial condition for leptogenesis, which we will explore in the next section.

\subsection{U(1)$_{L_\mu-L_\tau}$ gauge field generation under the slow-roll approximation}

We consider the axion inflation~\cite{Turner:1987bw,Freese:1990rb,Garretson:1992vt}\footnote{
    We call the inflaton in our model as the axion-like particle in the sense that the interaction term respects the softly-broken shift symmetry and parity and that its potential is close to flat.
}, in which a pseudo-scalar inflaton $\phi$ couples to the U(1)$_{L_\mu-L_\tau}$ gauge field $X_\mu$ through the Chern--Simons term. The action of our model is
\begin{align}
    S &= \int \mathrm{d}^4x\,\sqrt{-g}\,(\mathcal{L}_{\mathrm{inf}}+\mathcal{L}_{\mathrm{SM}_{L_\mu-L_\tau}})\,,\notag\\
    \mathcal{L}_{\mathrm{inf}}&= \frac{1}{2}\partial_\mu\phi\partial^\mu\phi - V(\phi) - \frac{1}{4}X_{\mu\nu}X^{\mu\nu} - \frac{g_{\phi XX}}{4}\phi X_{\mu\nu}\tilde{X}^{\mu\nu}\,,
\end{align}
where $\mathcal{L}_{\mathrm{inf}}$ is the relevant part during the inflation, and $\mathcal{L}_{\mathrm{SM}_{L_\mu-L_\tau}}$ is the Lagrangian density for the SM fields gauged by the U(1)$_{L_\mu-L_\tau}$ symmetry. 
We assume that the particles in the SM are effectively absent during inflation and neglect the latter term in this section.
In the expression of $\mathcal{L}_{\mathrm{inf}}$, $V(\phi)$ is the inflaton potential, $X_{\mu\nu}\equiv\partial_\mu X_\nu-\partial_\nu X_\mu$ is the field strength of the U(1)$_{L_\mu-L_\tau}$ gauge field, $\tilde{X}^{\mu\nu}\equiv\frac{1}{2}\epsilon^{\mu\nu\rho\sigma}X_{\rho\sigma}$ with the convention $\epsilon^{0123}=+1/\sqrt{-g}$ is the dual of the field strength tensor. Note that the coupling constant $g_{\phi XX}$ has mass dimension $-1$.

During the slow-roll regime of the inflation, the Hubble parameter $H_{\rm inf}\simeq (V/3M_\mathrm{Pl}^2)^{1/2}$ with $M_\mathrm{Pl}$ being the reduced Planck mass and the inflaton velocity $\dot\phi\simeq -V_{,\phi}/3H_{\mathrm{inf}}$ are approximately constant. On top of this background, the equation of motion for the U(1)$_{L_\mu-L_\tau}$ gauge field is~\cite{Anber:2009ua}
\begin{align}
    \big[\partial_\eta^2 + k(k \mp 2\xi a H_{\rm inf})\big] X_\pm(\eta,\vec k)=0\,,\quad
    \eta=-\dfrac{1}{aH_{\rm inf}}\,,
\end{align}
where we take the radiation gauge ($X_0=0$, $\vec k\cdot\vec X=0$), and $X_\pm$ are the right- and left- polarization modes so that $\vec X=\sum_\pm X_\pm\vec\epsilon_\pm$, $i\vec k\times\vec\epsilon_\pm=\pm k \vec\epsilon_\pm$. 
We have introduced the instability parameter $\xi\equiv g_{\phi XX}\dot{\phi}/(2H_{\rm inf})$, which triggers a tachyonic instability in one of the helicity modes $X_\pm$, depending on the sign of $\xi$.
 
Assuming the Bunch--Davies-like initial condition at the asymptotic past $k\eta\to-\infty$, the solution for the equation of motion in the slow-roll regime (constant $\xi$ approximation) is expressed using the Whittaker function \cite{Anber:2009ua,Jimenez:2017cdr}, in which one of the helicity modes $X_\pm(\eta,\vec k)$ for a particular wavelength range $0\lesssim -k\eta\lesssim 2\vert\xi\vert$ are amplified.
Consequently, at the end of the slow-roll inflation, the U(1)$_{L_\mu-L_\tau}$ gauge field is known to have the following stationary attractor properties for $\vert\xi\vert\gtrsim 4$~\cite{Jimenez:2017cdr}:
\begin{align}
    \langle \vec E_{\rm phys}^2\rangle &\simeq 2.6\times10^{-4}\,\frac{e^{2\pi\vert\xi\vert}}{\vert\xi\vert^3} H_{\rm inf}^4\,, \\
    \langle \vec B_{\rm phys}^2\rangle &\simeq 3.0\times10^{-4}\,\frac{e^{2\pi\vert\xi\vert}}{\vert\xi\vert^5} H_{\rm inf}^4\,, \label{eq:B2inf}\\
    \label{eq:EdotB}
    \langle \vec E_{\rm phys}\cdot\vec B_{\rm phys}\rangle &\simeq  -{\rm{sgn}}(\xi)\cdot2.6\times10^{-4}\,\frac{e^{2\pi\vert\xi\vert}}{\vert\xi\vert^4} H_{\rm inf}^4\,,
\end{align}
where $\vec E_{\rm phys}= -\partial_\eta\vec X/a^2$ and $\vec B_{\rm phys} = \vec\nabla\times\vec X/a^2$ are the physical U(1)$_{L_\mu-L_\tau}$ electromagnetic fields.
The physical coherence length of the gauge field is estimated as
\begin{equation}
    \xi_\mathrm{M, inf} \simeq 3.0 \left(\frac{|\xi|}{5}\right) \frac{2 \pi }{H_\mathrm{inf}} \simeq 3.7 \frac{|\xi|}{H_\mathrm{inf}}\,. \label{eq:CohLengthInf}
\end{equation}
With the stationary solution for $\langle \vec E_{\rm phys}\cdot\vec B_{\rm phys}\rangle$,  which is homogeneous and isotropic across over the volume larger than the Hubble volume, we roughly estimate the magnetic helicity density 
$h_{\rm phys}\equiv\langle a^{-1}\vec X\cdot \vec B_{\rm phys}\rangle$ from Eq.~\eqref{eq:EdotB} as~\cite{Domcke:2022kfs,Fukuda:2024pkh}
\begin{align}
    \label{eq:hx_initial}
    h_{\rm phys} \simeq -\left.\frac{2\langle\vec E_{\rm phys}\cdot\vec B_{\rm phys}\rangle}{3H_{\rm inf}}\right|_{\rm end}
    \simeq 1.7\times10^{-4}\,\frac{e^{2\pi\xi}}{\xi^4} H_{\rm inf}^3\,.
\end{align}

In this estimate, we do not take into account the backreaction on the dynamics of the inflaton and cosmic expansion, as well as other fields in the system.  We suppose that U(1)$_{L_\mu-L_\tau}$ charged particles become sufficiently heavy such that  the Schwinger effect is suppressed.\footnote{
    We give large masses to the $L_\mu-L_\tau$-charged fermions through the Weinberg operator, which explicitly violates the lepton symmetry and hence anomaly equation~\cite{Hu:2025nep}.
    Therefore, finite $X\tilde X$ without lepton number generation is consistent.
} 
Thus, its backreaction on the gauge field amplification as well as on the cosmic expansion 
can be neglected. See Ref.~\cite{Hu:2025nep} for a concrete model to realize such a situation. 
On the other hand, the backreaction of the produced gauge fields on the inflaton dynamics as well as 
on the cosmic expansion still exists and gives an upper bound on $\xi$, depending on the Hubble parameter. 
The former condition
\begin{equation}
    \delta_F = \frac{\rho_{EE} + \rho_{BB}}{\rho_\mathrm{inf}} = \frac{\frac{1}{2} (\langle \vec E_{\rm phys}^2\rangle +\langle \vec B_{\rm phys}^2\rangle) }{3 H_\mathrm{inf}^2 M_\mathrm{Pl}^2} \ll 1,
    \label{eq:BRinf}
\end{equation}
is found to give a more stringent constraint, which in turn determines the upper bound of the Hubble parameter for a given $\xi$ and also the helicity density~\cite{Jimenez:2017cdr}. 

\subsection{U(1)$_{L_\mu-L_\tau}$ magnetic helicity as the initial condition for leptogenesis}
So far we have not discussed the $\mathcal{L}_{\mathrm{SM}_{L_\mu-L_\tau}}$ term, turning it off from the effective theory for the inflation dynamics.  We now turn it on, by supposing that the U(1)$_{L_\mu-L_\tau}$ charged fermions become light after inflation, for the evolution of the system afterwards.

Let us denote the gauge coupling to be $g_{L_\mu-L_\tau}$.  The anomaly equation then tells that, although the U(1)$_{L_\mu-L_\tau}$ symmetry is anomaly free, $\partial_\mu J^\mu_{L_\mu-L_\tau}=0,$ the total lepton number $L=L_e+L_\mu+L_\tau$ is violated~\cite{Foot:1990mn,He:1990pn} as
\begin{align}
    \partial_\mu J_L^\mu =2\partial_\mu J_{L_\mu}^\mu=2\partial_\mu J_{L_\tau}^\mu= \frac{g_{L_\mu-L_\tau}^2}{8\pi^2} X_{\mu\nu}\tilde X^{\mu\nu}\,  \quad \text{while} \quad \partial_\mu J^\mu_{L_e} = 0\,. \label{anomalyeq}
\end{align}
Integrating over the whole space, we understand that a particular combination of the comoving $B-L$ number $Q_{B-L}=\int {\rm d}^3x\,J_{B-L}^0$ (the difference between the baryon number and lepton number) and the comoving magnetic helicity $\mathcal{H}_X\equiv\int {\rm d}^3x\,\vec X\cdot \vec B_X$, namely, $Q_{B-L} + (g_{L_\mu-L_\tau}^2/4\pi^2)\mathcal{H}_X$ is conserved. 

The anomaly equation implies that, when the comoving magnetic helicity $\mathcal{H}_X$ decays, it sources the comoving lepton number $Q_L$ with the relation in terms of number density, 
\begin{equation}
    \Delta n_{L,\mathrm{phys}} = \frac{g_{L_\mu-L_\tau}^2}{4 \pi^2} \Delta h_\mathrm{phys}\,. \label{eq:conselaw}
\end{equation}
For later purpose, it is convenient to parametrize the magnetic helicity density 
normalized by the entropy density:
\begin{align}
    \label{eq:etaH_def}
    \eta_H\equiv\frac{g_{L_\mu-L_\tau}^2}{8\pi^2} \frac{h_{\rm phys}}{s_{\rm phys}}\,, 
\end{align}
which will be related to the lepton-to-entropy ratio
\begin{equation}
     \eta_H \sim
    -\frac{n_{L_\mu,\,{\rm phys}}}{s_{\rm phys}}
    =-\frac{n_{L_\tau,\,{\rm phys}}}{s_{\rm phys}} = - \frac{1}{2} \eta_L^{\rm tot}\,,
    \label{eq:equivalentlepton}
\end{equation}
a convenient parameter to compare with observations. 
Here $\sim$ becomes an equality if we suppose that the magnetic helicity fully converts into the lepton number. 
$s_{\rm phys}$ is the physical entropy density,
\begin{align}
    s_{\rm phys} = \frac{2\pi^2}{45} g_{*s} T_{\rm reh}^3 = \frac{2\pi^2}{45} g_{*s}\left(\frac{90}{\pi^2 g_*}\right)^{3/4} (H_{\rm inf}M_{\rm Pl})^{3/2}\,, 
\end{align}
supposing the instantaneous reheating, 
$T_\mathrm{reh} = (90/\pi^2 g_*)^{1/4} (H_\mathrm{inf} M_\mathrm{Pl})^{1/2}$.
Here $g_*$ and $g_{*s}$ are the effective relativistic degrees of freedom for energy density and entropy, respectively, and $T_\mathrm{reh}$ is the reheating temperature. While in the minimal setup where
the Schwinger effect --- fermion pair production in the strong electric field --- is active, the resultant helicity-to-entropy ratio has an upper bound,
$\eta_H \lesssim 10^{-6}$~\cite{Domcke:2018eki,Hu:2025nep}.
By supposing the Schwinger effect is turned off during inflation, the 
helicity-to-entropy ratio corresponding to $\eta_H \sim 10^{-3}$ becomes viable without violating the energy bound Eq.~\eqref{eq:BRinf}) for $H_\mathrm{inf} \lesssim 10^{-2}$ GeV~\cite{Hu:2025nep}.

In our model, Eq.~\eqref{eq:hx_initial} is rewritten as
\begin{align}
    \label{eq:etaH_explicit}
    \eta_H = 9.4\times10^{-7}\, g_*^{-1/4} g_{L_\mu-L_\tau}^2\, \frac{e^{2\pi\vert\xi\vert}}{\xi^4}\left(\frac{H_{\rm inf}}{M_{\rm Pl}}\right)^{3/2}\,
\end{align}
which we take as the initial condition for the post-inflationary conversion of magnetic helicity to the lepton number.
By suppressing the Schwinger effect, we take
zero initial lepton asymmetry, $\eta_{L_e} = \eta_{L_\mu}=\eta_{L_\tau}=0$.
We will consider the MHD decay of the magnetic helicity and, accordingly, the generation of baryon and lepton numbers in the following sections.
\section{Baryogenesis from decaying U(1)$_{L_\mu-L_\tau}$ magnetic helicity}
\label{Sec:Baryogenesis}

\begin{figure}[htbp]
\centering
\includegraphics[width=1\textwidth]{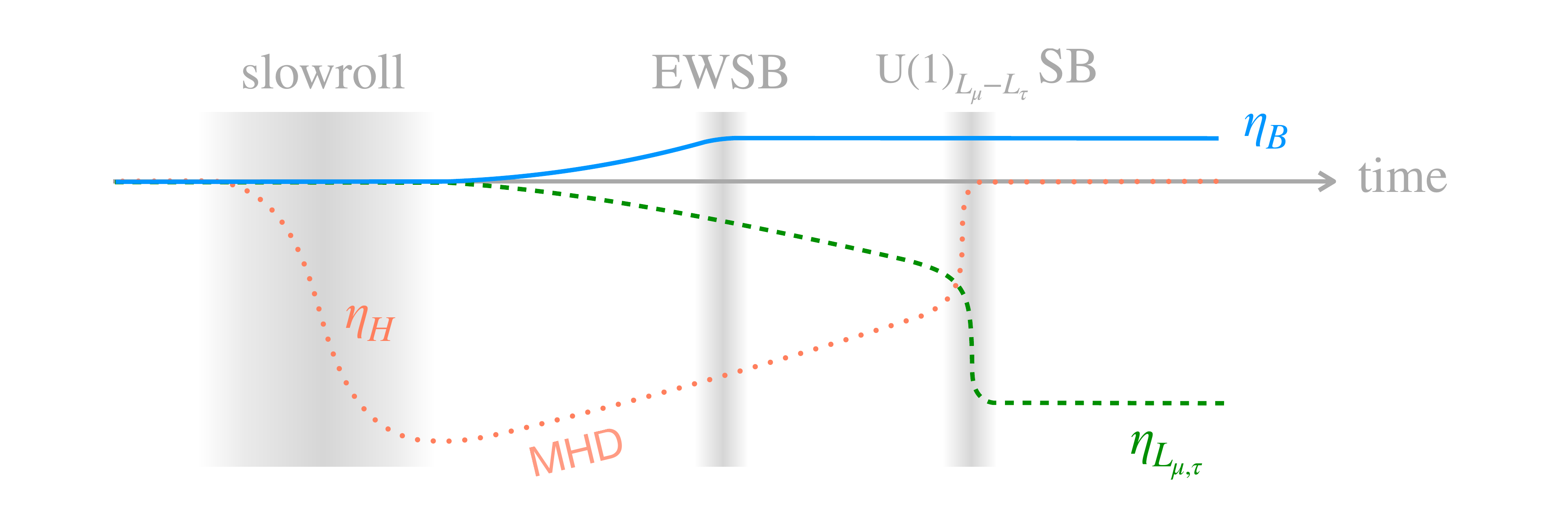}
    \caption{
    A schematic figure representing our scenario. We show the evolution of the $\mathrm{U}(1)_{L_{\mu}-L_\tau}$ magnetic helicity $\eta_H$ in the red-dotted line, the lepton asymmetry $\eta_{L_\mu}$ and $\eta_{L_\tau}$ in the green dashed line, and the baryon asymmetry $\eta_B$ in the blue solid line. In the slow-roll phase in the inflation, the magnetic helicity is generated through the $\phi X\tilde X$ coupling, and end up with Eq.~\eqref{eq:etaH_explicit}. We assume $\eta_{L_{\mu,\tau}}=\eta_B=0$ at the end of inflation. Afterwards, the magnetic helicity is dissipated in the MHD evolution, which is associated with the lepton asymmetry generation. Before the electroweak symmetry breaking, the electroweak sphaleron is effective and washes out $B+L$, implying the generation of $\eta_B$. Finally, at the U(1)$_{L_\mu-L_\tau}$ symmetry breaking, the magnetic helicity decays completely and the lepton asymmetry is fixed after additional induction. From the observational side, the observations of cosmic microwave background (CMB) and primordial light elements determine the baryon asymmetry of the universe through the Big-Bang nucleosynthesis (BBN), and they also give an upper bound for a
    possible large lepton asymmetry, which constrains the parameter space of our scenario.
    }
\label{fig:evo_schematic}
\end{figure}

In this section, we discuss the decay of the U(1)$_{L_\mu-L_\tau}$ magnetic field and its helicity through the dissipation, which sources leptons and then baryons indirectly through the sphaleron processes 
(Fig.~\ref{fig:evo_schematic}). We apply the analytic description of the freely decaying magnetic fields, 
following Ref.~\cite{Uchida:2024ude} (Sec.~\ref{Sec:Evolution}) and treat the magnetic helicity decay as the background source in the Boltzmann equations for SM particles in the electroweak symmetric phase (Sec.~\ref{Sec:helicity_decay}) to discuss a possible baryogenesis scenario.
Here we consider the case where the U(1)$_{L_\mu-L_\tau}$ spontaneously breaks after the electroweak symmetry breaking, which is realized for, {\it e.g.,} $g_{L_\mu-L_\tau} \sim 10^{-4}$ and the gauge boson mass is around $10^{-2}$ GeV.

\subsection{Evolution of the U(1)$_{L_\mu-L_\tau}$ magnetic field}
\label{Sec:Evolution}

We first explain the evolution of the U(1)$_{L_\mu-L_\tau}$ magnetic field before the electroweak symmetry breaking.
We neglect the regime $0\leq\tau-\tau_{\rm reh}\lesssim\vert \tau_{\rm reh}\vert$, where $\tau_{\rm reh}\,(<0)$ is the conformal time at the end of the inflation, and assume that $\vert\tau_{\rm reh}\vert\ll\tau - \tau_{\rm reh}\simeq\tau\,(\leq\tau_\mathrm{EW})$ with $\tau_\mathrm{EW}$ being the conformal time at the electroweak symmetry breaking.

We normalize the scale factor at the end of the inflation to be unity, $a_{\rm inf}=1$, and work in the conformal unit, which is a convenient choice~\cite{Brandenburg:1996fc} because the U(1)$_{L_\mu-L_\tau}$ gauge field is conformally invariant as long as the massless approximation is valid, which is violated just before the symmetry breaking.
Correspondingly, we introduce these comoving variables:
$B\equiv a^2 B_{\rm phys}$, $\xi_\mathrm{M}\equiv \xi_{\rm M, phys}/a$, and $h_X\equiv a^3 h_{\rm phys}$.
In particular, the initial magnetic field is characterized by $B(\tau_{\rm reh})=B_{\rm phys,inf}$, $\xi_\mathrm{M}(\tau_{\rm reh})=\xi_{\rm M,phys, inf}$, and $h_X(\tau_{\rm reh})=h_{\rm phys, inf}$. 

The subsequent evolution of the U(1)$_{L_\mu-L_\tau}$ magnetic field is governed by U(1)$_{L_\mu-L_\tau}$ magnetohydrodynamics, which we refer to just as MHD in this paper, since it is coupled to highly conducting plasma consisting of $L_\mu$s and $L_\tau$s.
We assume the absence of U(1)$_Y$ magnetic field. The MHD evolution is then classified according to the U(1)$_{L_\mu-L_\tau}$ magnetic Reynolds number,
\begin{align}
    {\rm Re}_{\rm M}\equiv \frac{v\xi_{\rm M}\sigma_X}{2\pi}\,, \label{Rem}
\end{align}
at reheating
where $v$ is the typical fluid velocity, $\xi_\mathrm{M}$ is the coherence length of the magnetic field, and $\sigma_X$ is the comoving U(1)$_{L_\mu-L_\tau}$ electric conductivity for the U(1)$_{L_\mu-L_\tau}$ current. 
We use
\begin{equation}
    \sigma_X =  \frac{a(T)g_{L_\mu - L_\tau}^2 T}{g'^4}
\end{equation} 
by adopting the Drude model~\cite{Arnold:2000dr,Kamada:2018kyi,Fukuda:2024pkh}.
Here $g'$ is the  gauge coupling for the hyper U(1) gauge interaction, $g' =  0.35$. 
In the following, we neglect the CME contribution when we determine the evolution of $B$ and $\xi_{\rm M}$. 
This is because, in the parameter space of our interest, chiral asymmetry is sufficiently small 
and the magnitude of the CME term relative to that of the Ohmic term in the induction equation remains below a few percent level 
such that the chiral plasma instability~\cite{Joyce:1997uy,Akamatsu:2013pjd} does not have a significant impact, see also Refs.~\cite{Brandenburg:2023rul,Brandenburg:2023aco}.  
\subsubsection{Non-linear turbulent case}
If ${\rm Re}_{\rm M,reh}\gg1$, advection can dominate over Ohmic diffusion.
The U(1)$_{L_\mu-L_\tau}$ magnetic field may be initially frozen\footnote{
    Dissipation effects at smaller scales are negligible at larger scales.
}, while the kinetic energy present at reheating stays comparable to the U(1)$_{L_\mu-L_\tau}$ magnetic energy.
Just after the frozen regime, the U(1)$_{L_\mu-L_\tau}$ magnetic field non-linearly decays in a transient equipartition regime.
Finally, the kinetic energy becomes subdominant, and magnetically dominated turbulence transports the U(1)$_{L_\mu-L_\tau}$ magnetic energy down to dissipation scales~\cite{Uchida:2024ude}.

\paragraph{Frozen regime.}
We estimate the initial fluid velocity at reheating by assuming an approximate equipartition between the magnetic and kinetic energies, $v_{\rm reh} \sim B_{\rm inf}/(a_{\rm reh}T_{\rm reh})^2$~\cite{Banerjee:2004df,Domcke:2019mnd,Fukuda:2024pkh} 
and approximately the same coherence length for kinetic and U(1)$_{L_\mu-L_\tau}$ magnetic fields, given in Eq.~\eqref{eq:CohLengthInf}.
In the equipartition regime, the only timescale that is characteristic to the system is the eddy turnover time
\begin{align}
    \tau_{\rm eddy, reh} \equiv \dfrac{\xi_{\rm M, reh}}{2 \pi v_{\rm reh}}\sim\dfrac{(a_{\rm reh}T_{\rm reh})^2\xi_{\rm M, inf}}{2 \pi B_{\rm inf}}\,.
\end{align}
In the frozen regime, the conformal time $\tau$ is smaller than the eddy turnover time, and the fields cannot decay. In particular, $B$ and $\xi_{\rm M}$ remain approximately constant~\cite{1993noma.book.....B,Uchida:2024ude}.

This regime terminates when the conformal time catches up the eddy turnover time,
\begin{align}
    \label{eq:cond-to-determine-xeq}
    \tau =\tau_{\rm eddy,reh} \simeq \dfrac{0.6\,\vert\xi\vert T_{\rm reh}^2}{H_{\rm inf}B(\tau_{\rm reh})}\,, 
\end{align}
where $B(\tau_{\rm reh})$ will be evaluated using Eq.~\eqref{eq:B2inf}.

\paragraph{Equipartition regime.}

In the equipartition regime after the frozen regime, an analytic description of the kinetic and U(1)$_{L_\mu-L_\tau}$ magnetic energy decay is possible\footnote{
    At the best of our knowledge, numerical investigations in this topic have not well been explored.
} by assuming two approximately conserved quantities~\cite{Hosking:2020wom} (regime~13 of Ref.~\cite{Uchida:2024ude})
: the U(1)$_{L_\mu-L_\tau}$ magnetic helicity $h_X \sim B^2 \xi_{\rm M}$ and the Saffman cross helicity integral $I_{\rm C} \sim B^2 v^2 \xi_{\rm M}^3$. 
Their conservation and the balance between the conformal time and the eddy turnover time, $\tau \simeq\tau_{\rm eddy}$, during a scaling regime implies
\begin{align}
    \label{eq:equi-scaling}
    \xi_{\rm M} \propto \tau^{1/2}\,,\qquad 
    B \propto \tau^{-1/4}\,,\qquad 
    v \propto \tau^{-1/2}\,,
\end{align}
or, equivalently, $\xi_{\rm M} \propto T^{-1/2}$, $B \propto T^{1/4}$.

This regime terminates when the energy density ratio $\Gamma \equiv (\rho+p)\,v^2/B^2$ drops well below unity~\cite{Hosking:2020wom,Uchida:2024ude}.
The ratio decays as $\Gamma \propto \tau^{-1/2}$, and if we assume the boundary between the equipartition and the subsequent magnetically dominated regimes at $\Gamma\sim 0.1$, the termination of this regime is
\begin{align}
    \label{eq:cond-to-determine-xmd}
    \tau\sim 10^2\,\tau_{\rm eddy, reh}\,,
\end{align}
when
\begin{align}
    \xi_{\rm M}\sim 10\,\xi_{\rm M,inf}\,,\quad
    B\sim 10^{-1/2}B(\tau_{\rm reh})\,,\quad
    {\rm Re}_{\rm M}\sim10^{1/2}\,{\rm Re}_{\rm M,reh}\,. \label{eq:EndEqPart}
\end{align}

\paragraph{Magnetically dominated turbulence.}
Since the U(1)$_{L_\mu-L_\tau}$ magnetic Reynolds number increases during the equipartition regime (Eq.~\eqref{eq:EndEqPart}), a magnetically dominated nonlinear regime follows after the equipartition regime.
In such nonlinear regimes, the Ohmic dissipation is negligible and the U(1)$_{L_\mu-L_\tau}$ magnetic helicity $h_X\sim B^2 \xi_{\rm M}$ stays approximately constant.
The nonlinear transport of the U(1)$_{L_\mu-L_\tau}$ magnetic energy down to smaller scales has not been understood well, but can be modeled by theories of magnetic reconnection \cite{Bhat:2020anu,Hosking:2020wom,Anandavijayan:2026ocw}.
We adopt the classification of the regimes in Ref.~\cite{Hosking:2022umv}, and introduce the Lundquist number, which is related to the U(1)$_{L_\mu-L_\tau}$ magnetic Reynolds number as $S = {\rm Re}_{\rm M}^{4/5}$ in the reconnection regimes \cite{Uchida:2024ude}.
An empirical value $S_{\rm c}\sim 10^4$ for the critical Lundquist number is known in the literature \cite{2011PhPl...18k1207J}.
A small Lundquist number $S\ll S_{\rm c}$ corresponds to the Sweet--Parker reconnection regime (regime~1 of Ref.~\cite{Uchida:2024ude}), where
\begin{align}
    \label{eq:scaling-SP}
    \xi_{\rm M}\propto \tau^{4/7}\,,\qquad B \propto \tau^{-2/7}\,.
\end{align}
A large Lundquist number $S\gg S_c$, on the other hand, corresponds to the fast reconnection regime (regime~5 of Ref.~\cite{Uchida:2024ude}), where
\begin{align}
    \label{eq:scaling-fast}
    \xi_{\rm M} \propto \tau^{2/3}\,,\qquad B \propto \tau^{-1/3}\,.
\end{align}

\subsubsection{Dissipation-dominated case}
If ${\rm Re}_{\rm M,reh}\ll1$, advection is subdominant, and the Ohmic diffusion directly dissipates the U(1)$_{L_\mu-L_\tau}$ magnetic energy.
The evolution of the U(1)$_{L_\mu-L_\tau}$ magnetic field reduces to the diffusion equation, which is easy to integrate formally and evaluate numerically.

Here we present an analytic simplification for an intuitive understanding, although we will carry out slightly more precise calculation in the next section.
Assuming that the magnetic field has never experienced nonlinear decay, we approximate that the strongly-peaked magnetic field spectrum generated at $\tau_{\rm reh}$ does not transform to the typical broken-power energy spectrum in a turbulence.
Then, the magnetic coherence length approximately stays constant until dissipation becomes effective. It becomes effective at
the timescale for the Ohmic dissipation,
\begin{align}
    \tau_{\rm Ohm} \equiv \frac{\xi_{\rm M}^2\sigma_X}{4\pi^2}
    \simeq \frac{\xi_{\rm M,inf}^2\sigma_X}{4\pi^2}\,. \label{eq:tauohm}
\end{align}
With this time scale, the magnetic-field amplitude would decay as fast as
\begin{align}
    B\sim\exp(-\tau/\tau_{\rm Ohm})\,, 
\end{align}
where $\sigma_X$ is approximated to be constant.

\subsection{U(1)$_{L_\mu-L_\tau}$ helicity decay as the source of leptons}
\label{Sec:helicity_decay}

In Sec.~\ref{Sec:Evolution}, we explained the MHD scaling laws, that determine $B$ and $\xi_{\rm M}$ as functions of time.
On top of it, we here derive the helicity dissipation rate, which serves as the source term for the lepton and baryon asymmetry generation.
Here we take into account the chiral magnetic effect, 
since it may backreact to the helicity dissipation rate.

The helicity decay rate for a maximally helical magnetic field including the CME in the MHD approximation is written as~\cite{Kamada:2016cnb,Kamada:2016eeb}
\begin{align}
    \label{eq:dhdx_general}
    \frac{{\rm d}h_X}{{\rm d}x} =-\dfrac{2}{T_{\rm reh}\sigma_X}\vec j_\mathrm{Ohm}\cdot\vec B
    = -\frac{4\pi}{T_{\rm reh}\sigma_X} \frac{B^2}{\xi_{\rm M}} +\frac{2\alpha_X}{\pi T_{\rm reh}\sigma_X}\mu_{5} B^2\,. 
\end{align}
Here $\mu_5$ denotes the comoving chiral chemical potential, 
\begin{equation}
    \mu_5 \equiv \sum_i \epsilon_i g_i Y_i^2 \mu_i\,,
    \label{eq:mu5}
\end{equation}
where $\epsilon_i = 1$ for right-handed fermions and $-1$ for left-handed fermions, $g_i$ is the internal gauge degree of freedom, 
$Y_i$ is the U(1)$_{L_\mu-L_\tau}$ charge, and $\mu_i$ is the comoving chemical potential  with the index $i$ labeling the fermion representations.
$\alpha_X\equiv g_{L_\mu-L_\tau}^2/(4\pi)$ is the U(1)$_{L_\mu-L_\tau}$ fine-structure constant. We adopted the magnetostatic approximation, namely $\vec j_\mathrm{Ohm} =\sigma_X( \vec E + \vec v \times \vec B) =\vec\nabla\times\vec B-(1/\pi)\alpha_X\mu_5\vec B$.
For convenience, we have introduced a dimensionless time variable
\begin{align}
    \label{vartaux}
    x \equiv\sqrt{\frac{90}{\pi^2 g_*}} \frac{M_{\rm Pl}}{T}\left(=\dfrac{T}{H}\right)\,,
\end{align}
so that the conformal time is given as a function of $x$, $\tau(x)=x/T_{\rm reh}$ with our normalization of the scale factor $a_{\rm reh}=1$. In what follows, we will find an expression of ${\rm d}h_X/{\rm d}x$ in terms of $B_{\rm inf}$ and $\xi_{\rm M,inf}$ in each regime of the magnetic field evolution.

\subsubsection{Helicity decay when ${\rm Re}_{\rm M,reh}\gg 1$}
\label{sec:NL_heldec}

\paragraph{Frozen regime.}
In the first epoch ($x_{\rm reh}<x<x_{\rm eq}$), the magnetic field remain approximately constant, $B(x)=B_{\rm inf}$ and $\xi_{\rm  M}(x)=\xi_{\rm M, inf}$.
Therefore
\begin{align}
    \frac{{\rm d}h_X}{{\rm d}x}=
    -\frac{4\pi}{T_{\rm reh}} \frac{B_{\rm inf}^2}{\xi_{\rm M, inf}} \frac{1}{\sigma_X}
    + \frac{2\alpha_X}{\pi T_{\rm reh}\,\sigma_X} \mu_5(x) B_{\rm inf}^2\,.
\label{eq:dhdx_frozen}
\end{align}
This regime terminates at $x=x_{\rm eq}$ when the condition \eqref{eq:cond-to-determine-xeq} is satisfied, namely
\begin{align}
    x_{\rm eq}=T_{\rm reh}\tau_{\rm eddy,reh}\,.
\end{align}

\paragraph{Equipartition regime.}
In the second epoch ($x_{\rm eq}<x<x_{\rm md}$), the kinetic and magnetic energies are comparable, and the MHD scaling \eqref{eq:equi-scaling}, $B(x)=B_{\rm inf}(x_{\rm eq}/x)^{1/4}$, $\xi_{\rm M}(x)=\xi_{\rm M,inf}(x/x_{\rm eq})^{1/2}$ implies
\begin{align}
    \label{eq:dhdx_equip}
    \frac{{\rm d}h_X}{{\rm d}x}=
    -\frac{4\pi}{T_{\rm reh}} \frac{B_{\rm inf}^2}{\xi_{\rm M, inf}} \frac{1}{\sigma_X} \frac{x_{\rm eq}}{x}
    + \frac{2\alpha_X}{\pi T_{\rm reh}\,\sigma_X} \mu_5(x) B_{\rm inf}^2
    \left(\frac{x_{\rm eq}}{x}\right)^{1/2}\,.
\end{align}
This regime terminates at $x=x_{\rm md}$ when the condition \eqref{eq:cond-to-determine-xmd} is satisfied, namely
\begin{align}
    x_{\rm md}\sim10^2T_{\rm reh}\tau_{\rm eddy,reh}\,.
\end{align}

\paragraph{Magnetically dominated turbulence.}
In the final epoch ($x>x_{\rm md}$), the evolution of the magnetic field is different depending on the Lundquist number $S(x)$. By defining $B_{\rm md}\equiv B(x_{\rm md})$ and $\xi_{\rm M, md}\equiv\xi_{\rm M}(x_{\rm md})$, 
we obtain formulas for the fast-reconnection regime ($S(x_{\rm md})\gg S_{\rm c}$)
using Eq.~\eqref{eq:scaling-fast}, $B(x)=B_{\rm md}(x_{\rm md}/x)^{1/3}$, $\xi_{\rm M}(x)=\xi_{\rm M, md}(x/x_{\rm md})^{2/3}$:
\begin{align}
    \frac{{\rm d}h_X}{{\rm d}x}=
    -\frac{4\pi}{T_{\rm reh}} \frac{B_{\rm md}^2}{\xi_{\rm M, md}} \frac{1}{\sigma_X} \left(\frac{x_{\rm md}}{x}\right)^{4/3}
    + \frac{2\alpha_X}{\pi T_{\rm reh}\,\sigma_X} \mu_5(x) B_{\rm md}^2
    \left(\frac{x_{\rm md}}{x}\right)^{2/3}\,,
    \label{eq:dhdx_md_fast}
\end{align}
and the Sweet--Parker regime ($S(x_{\rm md})\ll S_{\rm c}$), 
using Eq.~\eqref{eq:scaling-SP}, $B(x)=B_{\rm md}(x_{\rm md}/x)^{2/7}$, $\xi_{\rm M}(x)=\xi_{\rm M, md}(x/x_{\rm md})^{4/7}$:
\begin{align}
    \frac{{\rm d}h_X}{{\rm d}x}=
    -\frac{4\pi}{T_{\rm reh}} \frac{B_{\rm md}^2}{\xi_{\rm M, md}} \frac{1}{\sigma_X} \left(\frac{x_{\rm md}}{x}\right)^{8/7}
    + \frac{2\alpha_X}{\pi T_{\rm reh}\,\sigma_X} \mu_5(x) B_{\rm md}^2
    \left(\frac{x_{\rm md}}{x}\right)^{4/7}\,.
    \label{eq:dhdx_md_SP}
\end{align}
At the transition points, $B$ and $\xi_{\rm M}$ are continuous, so the piecewise decay rate is continuous at the transition points.

\subsubsection{Helicity decay when ${\rm Re}_{\rm M,reh}\ll1$}

When ${\rm Re}_{\rm M}\ll1$, 
the evolution of the magnetic field is dominated by diffusion:
\begin{align}
    \frac{\partial \boldsymbol{B}}{\partial\tau} = \frac{1}{\sigma_X} \nabla^2 \boldsymbol{B}\,.
    \label{eq:diffusion_B}
\end{align}
The magnetic field from axion inflation is sharply peaked at the instability scale $\xi_{\rm M, inf}$; modes at larger scales are suppressed by $e^{-\pi\xi}$.  Consequently, diffusive damping does not significantly shift $\xi_{\rm M}$, which remains $\approx\xi_{\rm M, inf}$ throughout the ${\rm Re}_M<1$ stage.  Under this approximation, Eq.~\eqref{eq:diffusion_B} can be integrated directly from reheating,
\begin{align}
    \boldsymbol{B}(\tau) = \boldsymbol{B}(\tau_{\mathrm{reh}})\,
    \exp\!\Big[{-\frac{4\pi^2}{\xi_{\rm M, inf}^2\sigma_X} }(\tau-\tau_{\mathrm{reh}})\Big]\,.
\end{align}
Substituting this expression into the general Eq.~\eqref{eq:dhdx_general}, we obtain
\begin{align}
    \frac{{\rm d}h_X}{{\rm d}x} &= 
    -\frac{8\pi^2}{T_{\rm reh}\,\xi_{\rm M, inf}^2\,\sigma_X}\,h_X(x)
    + \frac{2\alpha_X}{\pi T_{\rm reh}\,\sigma_X}\,\mu_5(x)\,B(x)^2\nonumber\\
    &\simeq-\big[\frac{8\pi^2}{T_{\rm reh}\xi_{\rm M,inf}^2\sigma_X}h_{\mathrm{inf}}-\frac{2\alpha_X}{\pi T_{\rm reh}\sigma_X}\mu_{5}(x) B_{\mathrm{inf}}^2\big]\exp\big[-\frac{8\pi^2}{T_{\rm reh}\xi_{\rm M, inf}^2\sigma_X}(x-x_{\mathrm{reh}}) \big]\,.
\end{align}
Since $\sigma_X$ is almost constant, the magnetic fields are almost frozen at first and start to decay exponentially around $\tau \sim \tau_\mathrm{Ohm}$. 

\subsection{Kinetic transport among baryons and leptons}
In this section, we explain the evolution of baryon and lepton asymmetries sourced by the helicity decay.
We describe the system by a set of Boltzmann equations for SM particles, and approximate the solution by assuming chemical equilibrium.

\subsubsection{Full Boltzmann system}

In the temperature window $T_{\rm reh}>T>T_{\rm EW}\approx 140 \text{GeV}$\footnote{
    Numerically we will find that the baryon asymmetry is fixed at $T \simeq 140$ GeV in the non-linear turbulent case. 
}, 
three classes of processes affect the baryon/lepton asymmetries: (i) the helicity decay as a lepton-number
source $\gamma_X$; (ii) the electroweak ($\gamma_{\rm w}$) and strong ($\gamma_{\rm s}$) sphalerons~\cite{Klinkhamer:1984di,McLerran:1990de}, which non-perturbatively redistribute fermion asymmetry, with the former violating both lepton and baryon number;
(iii) the SM Yukawa interactions ($\gamma_e,\gamma_{u},\gamma_d$), which equilibrate fermions and Higgs chemical potentials.  

The Boltzmann equations for the present U(1)$_{L_\mu-L_\tau}$ model in the electroweak symmetric phase~\cite{Fujita:2016igl,Kamada:2016eeb} are given by 
\begin{align}
    \frac{\partial\eta_{Q_i}}{\partial x} &= -3\gamma_{\rm w}\sum_j\bigl(\eta_{Q_j}+\eta_{L_j}\bigr) 
    - 2\gamma_{\rm s}\sum_j\bigl(\eta_{Q_j}-\eta_{u_j}-\eta_{d_j}\bigr) \nonumber\\
    &\quad -\sum_j\gamma_{u}^{ij}\Bigl(\frac{\eta_{Q_i}}{6}+\frac{\eta_\varphi}{4}-\frac{\eta_{u_j}}{3}\Bigr)
    -\sum_j\gamma_{d}^{ij}\Bigl(\frac{\eta_{Q_i}}{6}-\frac{\eta_\varphi}{4}-\frac{\eta_{d_j}}{3}\Bigr)\,, \label{eq:etaQ}\\[4pt]
    \frac{\partial\eta_{L_i}}{\partial x} &= -2\gamma_X(\delta_{i\mu}+\delta_{i\tau})
    - \gamma_{\rm w}\sum_j\bigl(\eta_{Q_j}+\eta_{L_j}\bigr) 
    - \sum_j\gamma_{e}^{ij}\Bigl(\frac{\eta_{L_i}}{2}-\frac{\eta_\varphi}{4}-\eta_{e_j}\Bigr)\,, \label{eq:etaL}\\[4pt]
    \frac{\partial\eta_{u_i}}{\partial x} &= \gamma_{\rm s}\sum_j\bigl(\eta_{Q_j}-\eta_{u_j}-\eta_{d_j}\bigr)
    + \sum_j\gamma_{u}^{ji}\Bigl(\frac{\eta_{Q_j}}{6}+\frac{\eta_\varphi}{4}-\frac{\eta_{u_i}}{3}\Bigr)\,, \label{eq:etau}\\[4pt]
    \frac{\partial\eta_{d_i}}{\partial x} &= \gamma_{\rm s}\sum_j\bigl(\eta_{Q_j}-\eta_{u_j}-\eta_{d_j}\bigr)
    + \sum_j\gamma_{d}^{ji}\Bigl(\frac{\eta_{Q_j}}{6}-\frac{\eta_\varphi}{4}-\frac{\eta_{d_i}}{3}\Bigr)\,, \label{eq:etad}\\[4pt]
    \frac{\partial\eta_{e_i}}{\partial x} &= \gamma_X(\delta_{i\mu}+\delta_{i\tau}) 
    + \sum_j\gamma_{e}^{ji}\Bigl(\frac{\eta_{L_j}}{2}-\frac{\eta_\varphi}{4}-\eta_{e_i}\Bigr)\,, \label{eq:etae}\\[4pt]
    \frac{\partial\eta_\varphi}{\partial x} &= -\sum_{i,j}\gamma_{u}^{ij}\Bigl(\frac{\eta_{Q_i}}{6}+\frac{\eta_\varphi}{4}-\frac{\eta_{u_j}}{3}\Bigr)
    + \sum_{i,j}\gamma_{d}^{ij}\Bigl(\frac{\eta_{Q_i}}{6}-\frac{\eta_\varphi}{4}-\frac{\eta_{d_j}}{3}\Bigr)\notag\\
    &\quad\;+ \sum_{i,j}\gamma_{e}^{ij}\Bigl(\frac{\eta_{L_i}}{2}-\frac{\eta_\varphi}{4}-\eta_{e_j}\Bigr)\,, \label{eq:etaphi}
\end{align}
where $\eta_{Q_i},\eta_{u_i},\eta_{d_i},\eta_{L_i},\eta_{e_i},$ and $\eta_\varphi$ are the abundance of fields (the left-handed quarks, right-handed up and down quarks, left-handed leptons, right-handed charged leptons, and Higgs, respectively), namely, their homogeneous number density normalized by the entropy density.
The index $i=1,2,3$ labels their generations.
Each field has its internal degrees of freedom, which is counted as one for a real scalar and a Weyl fermion.
This factor gives a weight to each field in the Yukawa interaction terms, where the coefficients $\gamma_{u (ed)}^{ij} \sim 10^{-2}|y_{u (ed)}^{ij}|^2/8\pi$ with $y_{u (ed)}^{ij}$ being the Yukawa coupling matrices are the rates of Yukawa interactions in the SM.

The sphaleron rates are
~\cite{Kamada:2016eeb,Annala:2023jvr,Benso:2026csa}
\begin{align}
    \gamma_{\rm s} &= 108 \alpha_s^5 \\
   \gamma_{\rm w} &= \dfrac{3}{2}\cdot \left\{\begin{array}{ll}
    6.23 \times 10^{-7}\,, & T\gtrsim  159.6 \,\mathrm{GeV} \\
    \exp[-153.1  + 0.858 \,(T/\mathrm{GeV})]\,,& T\lesssim  159.6 \,\mathrm{GeV}
    \end{array} \right.\,.
\end{align}
The effective helicity source $\gamma_X$ is defined as
\begin{align}
    \gamma_X &\equiv- \frac{g_{L_\mu-L_\tau}^2}{8\pi^2}\frac{1}{s_{\rm reh}}\,\frac{{\rm d}h_X}{{\rm d}x}\equiv\gamma_{\rm Source}-\gamma_{\rm CME}\eta_5\,, 
    \label{eq:ewsphaleronrate}
\end{align}
where $\gamma_{\rm Source}$ is the source term from the helicity decay, and $\gamma_{\rm CME}$ parameterizes the CME washout term (the first and the second terms in Eq.~\eqref{eq:dhdx_general}, respectively).

In the advection-dominated regime (${\rm Re}_{\rm M} \gg 1$), the magnetic field evolves according to the MHD scaling laws in Sec.~\ref{sec:NL_heldec}, and we summarize the effective source term in the following piecewise expression:
\begin{align}
    &\gamma_X^{{\rm Re}_{\rm M}>1}\notag\\
    &= 
    \begin{cases}
    \displaystyle \frac{2\alpha_X }{s_{\rm reh}T_{\rm reh}\,\sigma_X}\frac{B_{\rm inf}^2}{\xi_{\rm M,inf}}
    - \frac{6\alpha_X^2}{\pi^2 T_{\rm reh}^3\,\sigma_X} B_{\rm inf}^2\,\eta_5(x)\,,&\text{Frozen},\\[7pt]
    \displaystyle\frac{2\alpha_X }{s_{\rm reh}T_{\rm reh}\,\sigma_X}\frac{B_{\rm inf}^2}{\xi_{\rm M, inf}}\frac{x_{\rm eq}}{x}
    - \frac{6\alpha_X^2}{\pi^2 T_{\rm reh}^3\,\sigma_X} B_{\rm inf}^2\,\eta_5(x)\bigl(\frac{x_{\rm eq}}{x}\bigr)^{1/2}\,, &\text{Equipartition},\\[7pt]
    \displaystyle \frac{2\alpha_X }{s_{\rm reh}T_{\rm reh}\,\sigma_X}\frac{B_{\rm md}^2}{\xi_{\rm M, md}}\bigl(\frac{x_{\rm md}}{x}\bigr)^{4/3}
    - \frac{6\alpha_X^2}{\pi^2 T_{\rm reh}^3\,\sigma_X} B_{\rm md}^2\,\eta_5(x)\bigl(\frac{x_{\rm md}}{x}\bigr)^{2/3}\,,&\text{MD,~$S(x_{\rm md})>S_{\rm c}$},\\[7pt]
    \displaystyle\frac{2\alpha_X }{s_{\rm reh}T_{\rm reh}\,\sigma_X}\frac{B_{\rm md}^2}{\xi_{\rm M, md}}\bigl(\frac{x_{\rm md}}{x}\bigr)^{8/7}
    - \frac{6\alpha_X^2}{\pi^2 T_{\rm reh}^3\,\sigma_X} B_{\rm md}^2\,\eta_5(x)\bigl(\frac{x_{\rm md}}{x}\bigr)^{4/7}\,, &\text{MD,~$S(x_{\rm md})<S_{\rm c}$},\\
    \end{cases} 
    \label{eq:gammaX_Rm>1}
\end{align}
with 
\begin{equation}
    \eta_5(x)\equiv\sum_{j=2,3}\bigl(\eta_{e_j}(x)-\eta_{L_j}(x)\bigr)=\frac{T_{\rm reh}^2}{6s_{\rm reh}}\mu_5(x)\,, 
    \label{eq:mu5eta5}
\end{equation} 
where $s_{\rm reh}=s(T_{\rm reh})$ for instantaneous reheating.

In the diffusion-dominated regime (${\rm Re}_{\rm M}<1$), the corresponding source term is
\begin{align}
    \gamma_X^{{\rm Re}_{\rm M}<1}(x)
    ={}&
    \left[
    \frac{4\pi \alpha_X}
    {s_{\rm reh}T_{\rm reh}\sigma_X}
    \frac{h_{\rm inf}}{\xi_{\rm M, inf}^2}
    -
    \frac{6\alpha_X^2}
    {\pi^2T_{\rm reh}^3\sigma_X}
    B_{\rm inf}^2\eta_5(x)
    \right]
    \exp\!\left[
    -\frac{8\pi^2}
    {T_{\rm reh}\xi_{\rm M, inf}^2\sigma_X}
    (x-x_{\mathrm{reh}})\right]\,.
    \label{eq:gammaX_Rm<1}
\end{align}
Numerical integration of Eqs.~\eqref{eq:etaQ} -- \eqref{eq:gammaX_Rm<1} from $x_{\rm reh}$ to 
sufficiently later time than
$x_{\rm EW}$ yields $\eta_B$ as a function of our model parameters, $H_{\rm inf}$, $\xi$, and $g_{L_\mu-L_\tau}$. 
We do not take into account the effect of the evolution of the Higgs expectation value around the electroweak symmetry breaking, but it would change the net baryon number only by factors of order unity~\cite{Khlebnikov:1988sr,Harvey:1990qw,Khlebnikov:1996vj,Laine:1999wv,Benso:2026csa}.
We show the numerical evolution of $\eta_B$ for five values of $H_{\rm inf}$ as the solid curves in Fig.~\ref{fig:numvsana}, 
which will be further examined in Sec.~\ref{sec:numerical}.

\subsubsection{Chemical equilibrium approximation}

Let us investigate the evolution of the particle asymmetries driven by the Boltzmann equations introduced in the previous section. Supposing that all the 
SM Yukawa interaction and 
sphaleron processes are in equilibrium, 
which is the case when each reaction rate is larger than the Hubble parameter\footnote{
    It is realized only at $T\lesssim 100$ TeV when the electron Yukawa interaction is in equilibrium~\cite{Campbell:1992jd}. Strictly speaking, at higher temperature, Yukawa interactions as well as the sphaleron processes gradually become out of equilibrium, and several approximate conserved charge appears~\cite{Buchmuller:2001sr,Domcke:2020kcp,Domcke:2020quw}.
}, 
the abundance of the fields are not independent but related.
This approximation makes it possible to reduce the Boltzmann equations to those for only one asymmetry in our setup.

First, the quark asymmetries become generation independent due to the off-diagonal component of the down-type Yukawa matrix, while retaining separate lepton flavors because the U(1)$_{L_\mu-L_\tau}$ source acts on the $\mu$ and $\tau$ sectors but not on the electron sector. 
Note that the lepton Yukawa interaction does not mix the asymmetries in different generations in the basis where the lepton Yukawa matrix is diagonalized. The Yukawa equilibrium conditions are then given by
\begin{align}
    \eta_u =\frac1{2}\eta_Q+\frac3{4}\eta_\varphi\,,\qquad
     \eta_d=\frac1{2}\eta_Q-\frac3{4}\eta_\varphi\,,\qquad 
    \eta_{e_i}=\frac1{2}\eta_{L_i}-\frac1{4}\eta_\varphi\,, 
    \label{eq:chem_eq_relations}
\end{align}
where the generation index $i$ for quarks is omitted, while the electroweak-sphaleron equilibrium requires
\begin{align}
    3\eta_Q+\sum_i\eta_{L_i}=0\,. 
    \label{eq:weak_sph_eq}
\end{align}
We also require the hypercharge neutrality,
\begin{align}
    0=3\times\left(\frac16\eta_Q+\frac23\eta_u-\frac13\eta_d\right)-\frac12\sum_i\eta_{L_i}-\sum_i\eta_{e_i}+\frac12\eta_\varphi\,.
    \label{eq:hypercharge_neutrality}
\end{align}
Note that the strong-sphaleron equilibrium condition is automatically satisfied.
Substituting the Yukawa relations into Eq.~\eqref{eq:hypercharge_neutrality} and eliminating $\sum_i\eta_{L_i}$ with Eq.~\eqref{eq:weak_sph_eq}, we 
find that the quark and Higgs abundances satisfy proportionality relations, 
\begin{align}
    \eta_\varphi=-\frac{8}{7}\eta_Q\,,\qquad 
    \eta_u=-\frac5{14} \eta_Q\,,\qquad 
    \eta_d=\frac{19}{14} \eta_Q\,.
    \label{eq:all_in_etaQ}
\end{align}
Consequently, the baryon asymmetry can be expressed in terms of the left-handed quark abundance as
\begin{align}
    \eta_B=\frac13\times3\left(\eta_Q+\eta_u+\eta_d\right)=2\eta_Q\,.
    \label{eq:etaB_etaQ}
\end{align}

Let us now turn to the lepton sector. We introduce the flavored $B-L$ asymmetries,
\begin{align}
    \Delta_i\equiv\frac{\eta_B}{3}-\left(\eta_{L_i}+\eta_{e_i}\right)\,,\qquad 
    i=e,\mu,\tau\,, 
    \label{eq:Delta_flavor}
\end{align}
which are conserved quantities in the absence of the $L_\mu-L_\tau$ gauge field. Note that the lepton Yukawa matrix can be diagonalized, and hence in the eigenbasis the asymmetries do not have to be identical for all flavors.
Using Eqs.~\eqref{eq:chem_eq_relations}, \eqref{eq:all_in_etaQ}, and \eqref{eq:etaB_etaQ},
the lepton asymmetry can be expressed in terms of the baryon and the flavored $B-L$ asymmetries as
\begin{align}
    \eta_{L_i}=\frac{8}{63} \eta_B-\frac{2}{3} \Delta_i\,, \quad 
    \eta_{e_i}=\frac{13}{63} \eta_B-\frac{1}{3} \Delta_i \,.
    \label{eq:lepton_flavor_eq}
\end{align}
Noting that the summation of the flavored $B-L$ asymmetries gives the total $B-L$ asymmetry, 
\begin{align}
    \sum_i\Delta_i=\eta_B-\eta_L^{\rm tot}\equiv\eta_{B-L}\,,
    \label{eq:Delta_sum}
\end{align}
we find from Eqs.~\eqref{eq:lepton_flavor_eq} and \eqref{eq:weak_sph_eq} that the baryon asymmetry is related to the $B-L$ asymmetry as 
\begin{align}
    \eta_{B-L}=\frac{79}{28}\eta_B\,.
    \label{eq:BminL}
\end{align}
It is evident from these relations that the each fermion asymmetry can be described by the linear combination 
of the flavored $B-L$ asymmetries, $\Delta_i$. 

It is straightforward to rewrite the Boltzmann equation in terms of those 
for the flavored $B-L$ asymmetries, $\Delta_i$.
The ${\rm U}(1)_{L_\mu-L_\tau}$ anomaly sources $\Delta_\mu$ and $\Delta_\tau$ equally, whereas $\Delta_e$ has no direct source. 
From the Boltzmann equations (Eqs.~\eqref{eq:etaQ} – \eqref{eq:etae}) we find that 
the Yukawa interactions and the strong and electroweak sphaleron terms are canceled, and consequently
we obtain simple evolution equations,
\begin{align}
    \frac{\mathrm{d}\Delta_e}{{\rm d}x}=0\,,\qquad
    \frac{\mathrm{d}\Delta_\mu}{{\rm d}x}=\frac{\mathrm{d}\Delta_\tau}{{\rm d}x}= \gamma_X\,.
    \label{eq:Delta_evolution}
\end{align}
For vanishing initial fermion asymmetries, the solution satisfies
\begin{align}
    \Delta_e=0\,,\qquad \Delta_\mu=\Delta_\tau\,.
    \label{eq:Delta_solution}
\end{align}
Together with Eqs.~\eqref{eq:Delta_sum} and \eqref{eq:BminL}, we obtain
\begin{align}
    \Delta_\mu=\Delta_\tau=\frac12\eta_{B-L}=\frac{79}{56}\eta_B\,, 
    \label{eq:Delta_mutau}
\end{align}
which suggests that the evolution of the system can be described only by one asymmetry, 
$\eta_B$. Namely, we obtain the Boltzmann equation for the baryon asymmetry as
\begin{equation}
    \frac{\mathrm{d}\eta_B}{\mathrm{d}x} = \frac{56}{79} \gamma_X\,.   
    \label{eq:boltzetaB1}
\end{equation}

In the source term $\gamma_X$, we need to have the information of the chiral chemical 
potential for the $L_\mu-L_\tau$ gauge interaction (Eq.~\eqref{eq:mu5}), 
which is related to the chirality-to-entropy ratio, $\eta_5$ (Eq.~\eqref{eq:mu5eta5}).
By noting that Eq.~\eqref{eq:lepton_flavor_eq} gives
\begin{align}
    \eta_{e_i}-\eta_{L_i}=\frac{5}{63} \eta_B+\frac{1}{3} \Delta_i\,,
\end{align}
we find that the chirality-to-entropy ratio is also given by $\eta_B$ as
\begin{align}
    \eta_5=\frac{277}{252}\eta_B\,.
    \label{eq:eta5_etaB_flavored}
\end{align}
Consequently, with the equilibrium approximation, from Eq.~\eqref{eq:boltzetaB1} the Boltzmann equations of the system 
is
are reduced to that for baryon asymmetry in a closed form,
\begin{align}
    \frac{{\mathrm d}\eta_B}{{\rm d}x}=\frac{56}{79}\gamma_{\rm Source}-\frac{554}{711}\gamma_{\rm CME}\eta_B\,, 
    \label{eq:detaB_closed}
\end{align}
where $\gamma_\mathrm{Source}$ and $\gamma_\mathrm{CME}$ are determined by the 
dynamics of the magnetic field, independent of the baryon asymmetry.

Let us now analytically evaluate the Boltzmann equation Eq.~\eqref{eq:detaB_closed}. 
For $\eta_B(x_{\rm reh})=0$, the formal solution is
\begin{align}
    \eta_B(x)
    =\frac{56}{79}\int_{x_{\rm reh}}^x{\rm d}x'\,\gamma_{\rm Source}(x')\exp\!\left[-\frac{554}{711}\int_{x'}^x{\rm d}x''\,\gamma_{\rm CME}(x'')\right]\,.
    \label{eq:etaB_formal}
\end{align}
It can be further simplified if the system remains in the frozen regime.  
If we approximate that $\gamma_{\rm Source}$ and $\gamma_{\rm CME}$ stay constant, we obtain
\begin{align}
    \eta_B(x)=\frac{252}{277}\frac{\gamma_{\rm Source}}{\gamma_{\rm CME}}\left[1-\exp\!\left(-\frac{554}{711}\gamma_{\rm CME}(x-x_{\rm reh})\right)\right]\,.
    \label{eq:etaB_frozen}
\end{align}

Furthermore, the CME term can be omitted for the following reasons. The ratio of the CME washout term to the helicity-decay source in $\gamma_X$ (Eq.~\eqref{eq:ewsphaleronrate}) is evaluated as
\begin{align}
    \epsilon_{\rm CME}
    \equiv\left|\frac{\gamma_{\rm CME}\eta_5}{\gamma_{\rm Source}}\right|
    =\frac{3\alpha_Xs_{\rm reh}\xi_{\rm M,\rm inf}}{\pi^2T_{\rm reh}^2}|\eta_5|=\frac{277}{1890}\alpha_Xg_*\left(T_{\rm reh}\xi_{\rm M,\rm inf}\right)|\eta_B|\,,
    \label{eq:CME_source_ratio}
\end{align}
where we used $s_{\rm reh}=2\pi^2g_*T_{\rm reh}^3/45$ and Eq.~\eqref{eq:eta5_etaB_flavored}. With $\xi_{\rm M,\rm inf}=3.7|\xi|/H_{\rm inf}$, $T_{\rm reh}=(90/\pi^2g_*)^{1/4}\sqrt{H_{\rm inf}M_{\rm Pl}}$, $\alpha_X=(3\times10^{-4})^2/(4\pi)$, and $g_*=106.75$, while taking $|\eta_B|=\eta_B^{\rm obs}$ as a reference value,  we obtain
\begin{align}
    \epsilon_{\rm CME}
    \simeq1.9\times10^{-17}|\xi|\sqrt{\frac{M_{\rm Pl}}{H_{\rm inf}}}\simeq3.0\times10^{-8}|\xi|\left(\frac{H_{\rm inf}}{1\,{\rm GeV}}\right)^{-1/2} \left(\frac{\eta_B}{\eta_B^\mathrm{obs}}\right)\,.
    \label{eq:CME_smallness}
\end{align}
We find that, for the parameter range considered here, with $|\xi|\sim10$ and $H_{\rm inf}\gtrsim10^{-10}\,{\rm GeV}$ 
while $\eta_B < \eta_B^\mathrm{obs}$ during the evolution, $\epsilon_{\rm CME}$ does not exceed a few percent. We therefore neglect the CME washout in the following analytic estimate. Taking $\gamma_{\rm CME}\to0$ in Eq.~\eqref{eq:etaB_frozen}, we obtain
\begin{align}
    \eta_B(x)=\frac{56}{79}\gamma_{\rm Source}(x-x_{\rm reh})\,,
    \label{eq:etaB_noCME}
\end{align}
which is independent of $g_{L_\mu-L_\tau}$ for given $H_{\rm inf}$ and $\xi$.
In both the non-linear turbulent and the dissipation-dominated cases, once the frozen regime ends, the $\gamma_\mathrm{Source}$ decays not slower than 
$x^{-1}$;
see Eqs.~\eqref{eq:gammaX_Rm>1} and \eqref{eq:gammaX_Rm<1}. 
Together with the approximation that $\gamma_\mathrm{CME}$ is negligible, we conclude that there is not significant generation of baryon 
asymmetry anymore after the end of frozen regime. 
Therefore, the final baryon asymmetry is approximated as
\begin{align}
    \eta_B=\frac{56}{79}\gamma_{\rm Source}({\bar x}-x_{\rm reh})\,, \quad 
    {\bar x} \equiv \mathrm{min} \{x_\mathrm{f}, x_\mathrm{EW}\}\,, 
    \label{eq:netetaB}
\end{align}
where $x_\mathrm{f}$ is the time when the frozen regime terminates; see Eqs.~\eqref{eq:cond-to-determine-xeq} and \eqref{eq:tauohm}. 

\begin{figure}[htbp]
\centering
\includegraphics[width=0.9\textwidth]{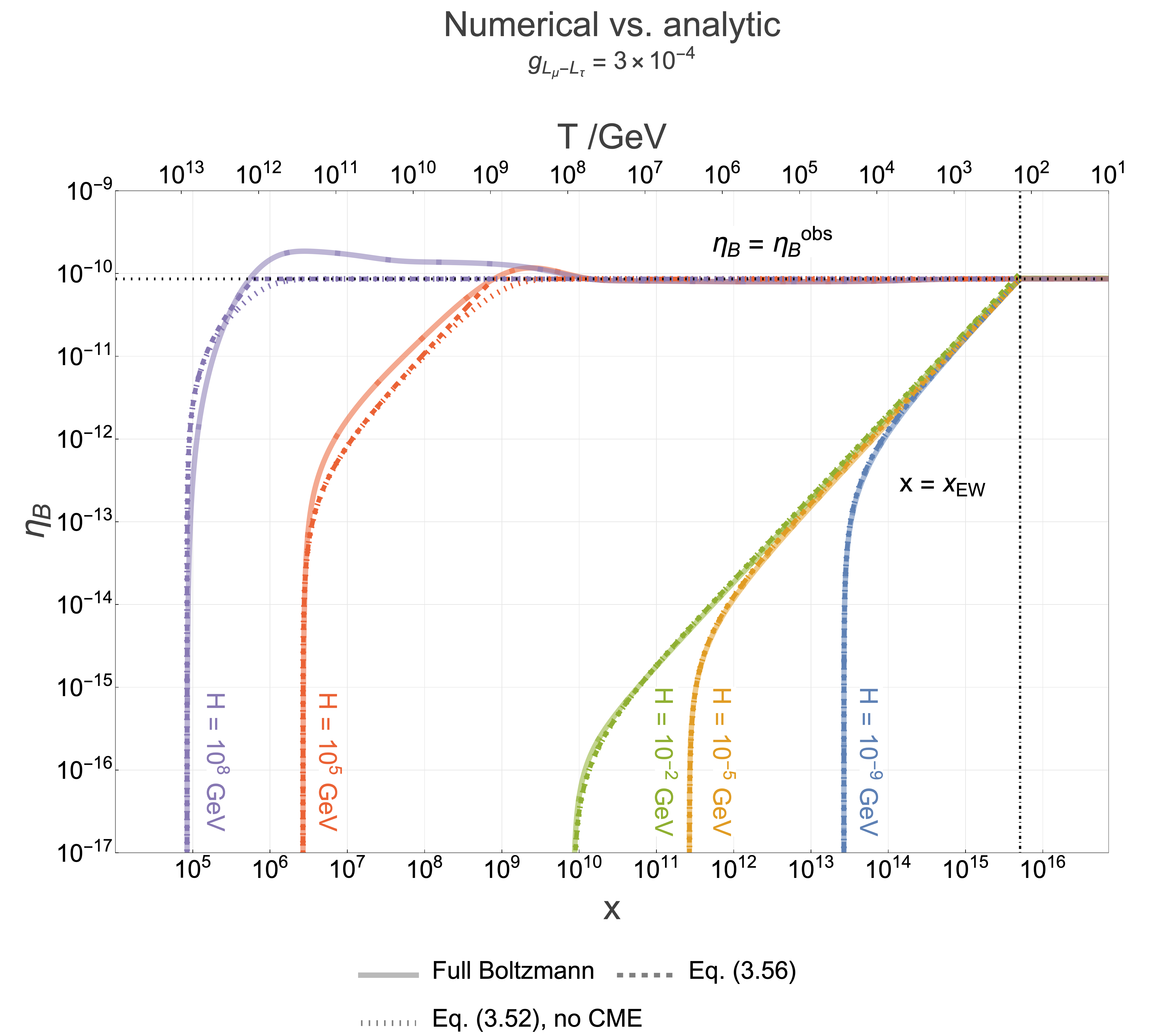}
\caption{
Comparison between the full numerical solution of the Boltzmann equations (solid lines, Eqs.~\eqref{eq:etaQ} – \eqref{eq:etaphi}) and the analytic approximations in the frozen regime for the baryon-number evolution with the chemical equilibrium conditions, neglecting the CME contribution.
The constant $\gamma_{\rm Source}$ approximation
( dashed lines, Eq.~\eqref{eq:etaB_noCME} for $x\leq {\bar x} $) as well as the 
solution taking into account the $\gamma_{\rm Source}$ evolution ( dotted lines, Eq.~\eqref{eq:etaB_formal}) are shown.  The blue, orange, green, red, and purple lines correspond to $H_{\rm inf}=10^{-9},10^{-5},10^{-2},10^{5},10^8\,{\rm GeV}$, respectively, while $\xi$ is determined so that the observed baryon asymmetry is obtained in the full numerical solutions and used for analytic approximations for given $H_\mathrm{inf}$.  
The horizontal dotted line denotes the observed baryon asymmetry, $\eta_B=\eta_B^{\rm obs}$, and the vertical dotted line marks the electroweak epoch, $x=x_{\rm EW}$.  
}
\label{fig:numvsana}
\end{figure}

Figure~\ref{fig:numvsana} compares the full numerical solutions of the Boltzmann equations.~\eqref{eq:etaQ} – \eqref{eq:etaphi} from $T=T_\mathrm{reh}$ down to 10 GeV
with the approximate formulae,
Eq.~\eqref{eq:etaB_noCME} with the frozen approximation and Eq.~\eqref{eq:etaB_formal} neglecting the CME contribution.
In the plot, we take $g_{L_\mu-L_\tau}$ to be $3\times 10^{-4}$.
We take representative values of $H_{\rm inf} = 10^{-9}, 10^{-5}, 10^{-2}, 10^5,$ and $10^8$ GeV.
The value of $\xi$ is chosen so that the observed baryon asymmetry is obtained by the full numerical solutions of the Boltzmann equations.
For given $H_\mathrm{inf}$, this value is used for plotting the results of approximated formulae. 
In the estimate of the final baryon asymmetry Eq.~\eqref{eq:netetaB},
${\bar x}$ is determined to be $x_\mathrm{EW}$ for 
the non-linear turbulent cases, $H_\mathrm{inf} = 10^{-5}$ and $10^{-9}$ GeV, and $x(\tau_\mathrm{Ohm}/2)$ for the dissipation-dominated case, $H_\mathrm{inf} = 10^5$ and $10^8$ GeV.\footnote{\samepage
    While the magnetic field strength decays roughly $\propto \exp[-\tau/\tau_\mathrm{Ohm}]$, magnetic helicity decays as $\propto \exp[-2 \tau/\tau_\mathrm{Ohm}]$.
    Thus we choose the end time of the frozen regime as $\tau_\mathrm{Ohm}/2$. 
} 
$H_\mathrm{inf}=10^{-2}$ GeV is somewhat at the intermediate regime between the non-linear turbulent and dissipation-dominated cases.

We see that the approximate formulae of Eqs.~\eqref{eq:etaB_noCME} and \eqref{eq:netetaB} 
describe the evolution of numerical results well. 
While the difference in the temperature evolution of the baryon asymmetry at an earlier time is relatively large, 
their prediction of the final baryon asymmetry to that of the full Boltzmann equation differs at most 10 \% level. 
This implies that the result is insensitive to the precise value of $g_{L_\mu-L_\tau}$ for given $H_{\rm inf}$ and $\xi$.
Analytic approximation gives a smaller value at $T<10^8$ GeV for  $H_\mathrm{inf}=10^8$ GeV and $10^5$ GeV, 
which can be understood as the violation of the equilibrium 
condition of the muon Yukawa interaction at $T\simeq 10^9$ GeV. 
Strictly speaking, in the non-linear turbulent cases, the baryon asymmetry is not completely fixed at $x=x_\mathrm{EW}$ and becomes constant only after the gradual freezeout of the electroweak sphaleron process, but the contribution during this period before reaching the constancy is negligibly small compared to the one at $x<x_\mathrm{EW}$. 
Consequently, Eq.~\eqref{eq:netetaB} gives a good approximated formula in predicting the baryon asymmetry of the Universe.

\subsection{Parameter space for the observed baryon asymmetry}
\label{sec:numerical}

\begin{figure}[htbp]
\centering
\includegraphics[scale=0.49]{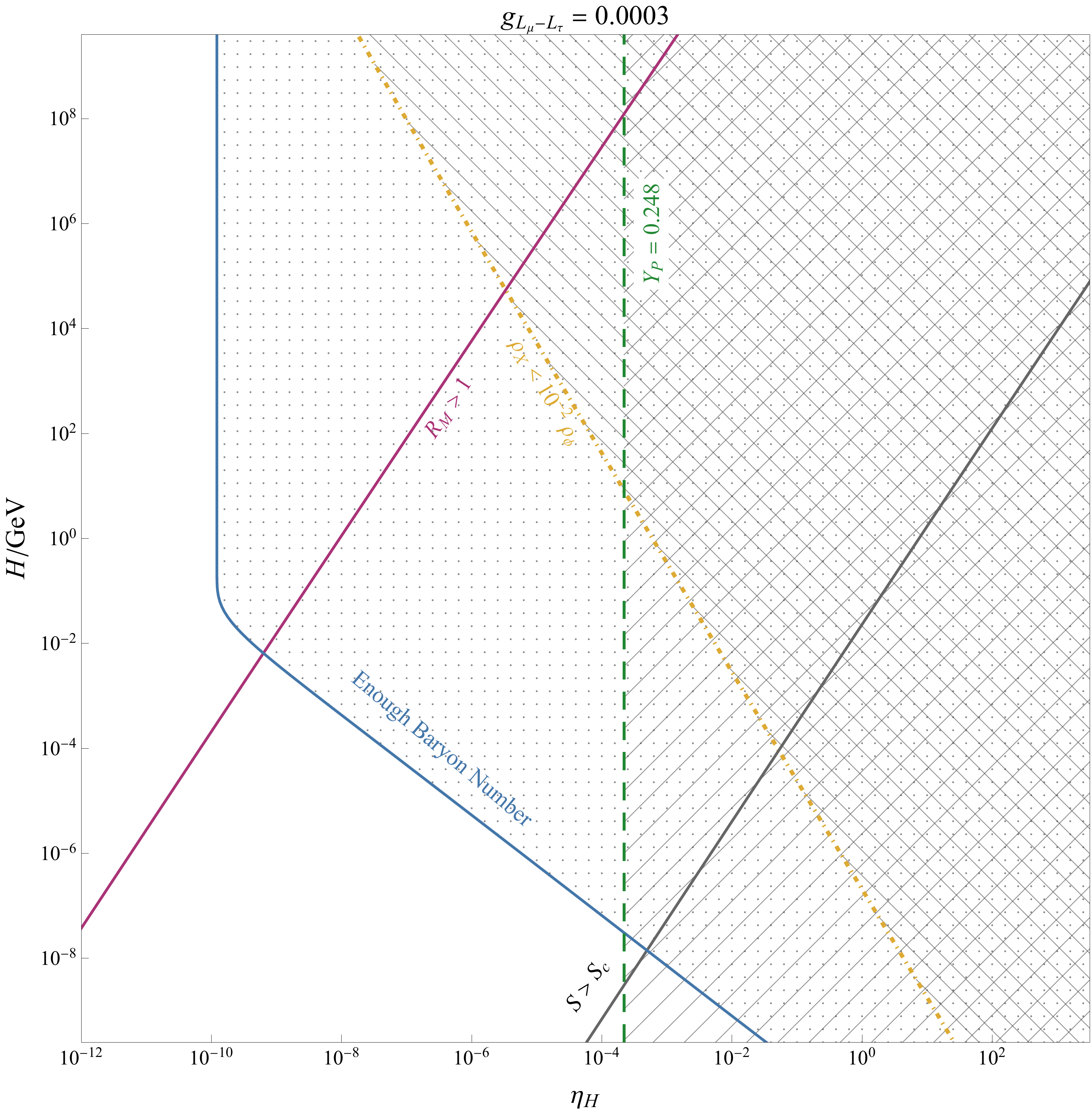}
\caption{
Parameter space in the $(\eta_H,H_{\rm inf})$ plane for $g_{L_\mu-L_\tau}=3\times10^{-4}$. The blue solid line shows the contour on which the 
the $L_\mu-L_\tau$ helicity dissipation generates the observed baryon asymmetry, $\eta_B=\eta_B^{\rm obs}$, through the electroweak sphaleron process. The gray stippled region, upper side of this contour is excluded due to the baryon overproduction. The purple solid line marks the ${\rm Re}_{\rm M}=1$ boundary, which determines if the $L_\mu-L_\tau$ magnetic field evolution is in the non-linear turbulent regime or dissipation-dominated regime. 
The dark-gray solid line denotes the $S=S_c$ boundary between the
Sweet--Parker and fast-reconnection regimes. 
This line does not change the behavior of the contour for the observed baryon asymmetry, since its net value is determined during the frozen regime, insensitive to the details of the magnetic reconnection mechanisms during the turbulence epoch.
The yellow dash-dotted line marks the $\rho_X=10^{-2}\rho_\phi$ boundary; see Ref.~\cite{Hu:2025nep} for the derivation of this boundary. The region above the line with downward-sloping diagonal hatching violates the gauge-field energy bound, $\rho_X<10^{-2}\rho_\phi$. 
The region with upward-sloping diagonal hatching, located to the right of the green dashed line, is excluded because the resulting lepton asymmetry leads to an overproduction of primordial helium, as evaluated using the numerical code of
Ref.~\cite{Domcke:2025jiy}. 
The green dashed vertical line indicates the boundary at $Y_P=Y_P^{\rm MAX}$ boundary, where we adopt $Y_P^{\rm MAX}=0.248$, corresponding to the $1\sigma$ upper limit of the Particle Data Group recommended value $Y_P=0.245\pm0.003$~\cite{ParticleDataGroup:2024cfk}.}
\label{parameterspace}
\end{figure}

In this section, we determine the parameter space that reproduces the observed baryon asymmetry and provide its physical interpretation. 
For this purpose, we need to solve the Boltzmann system~\eqref{eq:etaQ} --
\eqref{eq:gammaX_Rm<1}.  
As we have seen in the previous subsection, the baryon asymmetry is generated and fixed 
during the frozen regime of the magnetic field evolution. 
Here we improve the evaluation of the effective source term Eq.~\eqref{eq:gammaX_Rm>1} in the frozen regime, respecting the anomaly equation Eq.~\eqref{anomalyeq}.
By noting that
$\eta_H(x)=\eta_{H}(x_{\rm reh})-\frac{\eta_B(x)-\eta^{\rm tot}_L(x)}{2}$ follows from the conservation law Eq.~\eqref{eq:conselaw}, we use, instead of Eqs.~\eqref{eq:gammaX_Rm>1} and \eqref{eq:gammaX_Rm<1},
\begin{align}
    \gamma_X(x)&= \gamma_{\rm Source}(x)-\gamma_{\rm CME}(x)\eta_5(x)\nonumber \\
    &=\frac{8 \pi^2}{T_{\mathrm{reh}} \sigma_X \xi_{\rm M,inf}^2} \eta_H(x)-\frac{6 g_{L_\mu-L_\tau}^2 s_{\mathrm{reh}}}{\pi T_{\mathrm{reh}}^3 \sigma_X \xi_{\rm M,inf}} \eta_H(x) \eta_5(x) \,,
\end{align}
incorporating the decrease (even in the ``frozen'' regime) of the magnetic field strength due to the fermion generation.
With these treatments, we solve the Boltzmann equations and determine the contour $\eta_B=\eta_B^{\rm obs}$ in the $(\eta_H,H_{\rm inf})$ plane.  The result is shown as the solid blue line in Fig.~\ref{parameterspace}.  
Here we have chosen $g_{L_\mu-L_\tau}=3\times10^{-4}$. 

The contour for the observed baryon asymmetry passes through two regions of the parameter space. Namely, part of the contour lies in the ${\rm Re}_{\rm M}<1$ region, where the magnetic field evolution is governed entirely by the Ohmic diffusion, while  the rest lies in the ${\rm Re}_{\rm M}>1$ region, where
the magnetic field evolution is described by the non-linear turbulence.
The border of these two parts are shown in the purple solid line in Fig.~\ref{parameterspace}.
The analytic approximations obtained in Sec.~\ref{Sec:helicity_decay} provide a consistency check on the numerical integration.  Along the viable contour, we find that Eq.~\eqref{eq:netetaB}  reproduces the full numerical $\eta_B$ well.

The shape of the viable contour can be understood from the scaling of the initial magnetic field.  
In the dissipation-dominated case, magnetic helicity is
completely transferred to the $B-L$ asymmetry before the electroweak symmetry breaking. 
Therefore, the observed baryon asymmetry is realized for the vertical line (see Eq.~\eqref{eq:equivalentlepton}),
\begin{align}
    \eta_H = \dfrac{1}{2}\eta_{B-L}= \dfrac{79}{56} \eta_B^\mathrm{obs}\,.
\end{align}
In the non-linear turbulent regime, the contour runs diagonally across the $(\eta_H,H_{\mathrm{inf}})$ plane. This behavior can be understood from Eq.~\eqref{eq:netetaB}. Since the baryon asymmetry is mainly generated during the frozen regime, we approximately obtain
\begin{align}
    \eta_B\sim\gamma_{\rm Source}(x_{\rm EW}-x_{\rm reh})
    \sim \dfrac{\eta_H(x_{\rm EW}-x_{\rm reh})}{T_{\rm reh}\sigma_X\xi_{\rm M,inf}^2}
    \sim \dfrac{g'^4}{\alpha_X\xi^2}\eta_H\dfrac{H_{\rm inf}}{T_{\rm EW}}\,,
\end{align} 
neglecting the CME contribution; see Eq.~\eqref{eq:gammaX_Rm>1}. 
Since $\eta_H$ depends on $\xi$ exponentially (Eq.~\eqref{eq:etaH_explicit}),
the power-law dependence in $\xi$ of $\eta_B$ is negligible such that we can approximate $\eta_B \propto \eta_H H_\mathrm{inf}$. 
Consequently, the contour approximately follows $\eta_H\propto H_{\mathrm{inf}}^{-1}$.

In the $(\eta_H, H_\mathrm{inf})$ plane, some parameter regions are excluded. 
The first constraint is that the generated magnetic energy should be sufficiently smaller than that of the inflaton. 
The line with $\delta_F\ll 1$ (See Eq.~\eqref{eq:BRinf}) is shown in yellow dashed line in Fig.~\ref{parameterspace}. 
The second is the lepton overproduction constraint. 
In the nonlinear turbulent dominated regime, the magnetic field still exists after the electroweak symmetry breaking and generates a large lepton asymmetry at their complete decay, as has been studied in Ref.~\cite{Hu:2025nep} (but with the opposite sign). 
The flavored lepton asymmetries generated after electroweak symmetry breaking, $\eta_{L_\mu}$ and $\eta_{L_\tau}$, change the primordial 
helium abundance generated by the BBN. 
By evaluating the helium abundance $Y_P$ with the initial condition,  
$\eta_{L_\mu} = \eta_{L_\tau} =- \eta_H$, using the numerical code of
Ref.~\cite{Domcke:2025jiy}, 
we determine the upper bound of $\eta_H$, such that $Y_P < Y_P^\mathrm{MAX} = 0.248$, corresponding to the $1\sigma$ upper limit of the Particle Data Group recommended value $Y_P=0.245\pm0.003$~\cite{ParticleDataGroup:2024cfk}.
We find that, for the parameter region where baryons are not overproduced, the latter gives a stronger constraint
and the observed baryon asymmetry can be obtained for $\eta_H < 2.2  \times 10^{-4}$ and $H_\mathrm{inf} > 3.2 \times 10^{-8}$ GeV. 
Note that the net lepton asymmetry and baryon asymmetry have opposite signs; 
to explain the relatively small helium abundance suggested by the recent analysis of CMB~\cite{AtacamaCosmologyTelescope:2025blo} as well as 
the observation  of extremely metal poor galaxies~\cite{Yanagisawa:2025mgx}, which prefers a positive large lepton asymmetry in this scenario, we need to suppress the negative baryon asymmetry generated before the electroweak symmetry breaking through the earlier dissipation of magnetic helicity.

As has been indicated that in the previous subsection and Fig.~\ref{fig:numvsana}, baryon asymmetry is generated mostly
in the frozen stage. 
Therefore, the reconnection mechanisms which are determined by the Lundquist number are insensitive to the net baryon asymmetry.  
More importantly, it also has consequences for baryon isocurvature perturbation.  
The fact that the baryon asymmetry is generated before the onset of the cascade process means that the magnetic correlation length remains small 
at the time of baryon number generation.
Spatial inhomogeneities inherited from the inflationary production of helicity are then efficiently smoothed out by neutron diffusion before BBN, and hence the baryon isocurvature fluctuations are strongly suppressed~\cite{Kamada:2020bmb}.

\section{Discussion}
\label{Sec:Discussion}

We have studied the post-inflationary evolution of a helical U(1)$_{L_\mu-L_\tau}$ gauge field generated during axion inflation through the Chern-Simons interaction and its implications for baryogenesis.
As the magnetic field dissipates, the chiral anomaly generates flavored lepton asymmetries, which are partially converted
into baryon number through the electroweak sphaleron process.   
We describe the magnetic-field evolution using the MHD scaling laws 
explored in Ref.~\cite{Uchida:2024ude} and solve the coupled Boltzmann equations for the SM charges.  
Characterizing the inflationary magnetogenesis mechanism by the Hubble parameter during inflation $H_\mathrm{inf}$ and the helicity-to-entropy ratio just after inflation $\eta_H$, we find that the observed baryon abundance 
can be reproduced for $H_\mathrm{inf} > 3.2 \times 10^{-8}$ GeV and $\eta_H<2.2 \times 10^{-4} $, with the benchmark coupling $g_{L_\mu-L_\tau}=3\times10^{-4}$,  as shown by the blue contour in Fig.~\ref{parameterspace}. 
For ${\rm Re}_\mathrm{M}<1$, Ohmic diffusion governs the helicity dissipation
whereas for ${\rm Re}_\mathrm{M}>1$, the cascade process suppresses the helicity loss.

Since magnetic-field dissipation primarily generates a lepton asymmetry rather than a baryon asymmetry, only the helicity dissipated before the sphaleron freeze-out contributes to the final baryon abundance.  
Its magnitude depends both on the helicity generated during inflation and on the subsequent magnetic-field evolution.  
In the parameter space where the observed baryon asymmetry is reproduced, larger values of Hubble parameter correspond to $\mathrm{Re}_{\rm M}<1$, 
where the magnetic field  
dissipates efficiently, resulting in 
$|\eta_B| \sim |\eta_L| \sim |\eta_H|$. 
On the other hand, 
smaller values of Hubble parameter correspond to $\mathrm{Re}_{\rm M}>1$, where the magnetic field 
survives after the electroweak symmetry breaking, resulting in $|\eta_B| \ll |\eta_L| \sim |\eta_H|$.

In the latter case, we have assumed that the cascade process eventually terminates and magnetic helicity rapidly dissipates following the U(1)$_{L_\mu-L_\tau}$ symmetry breaking~\cite{Fukuda:2024pkh,Hu:2025nep}. 
If the symmetry breaking occurs after the electroweak symmetry breaking and before the BBN, the resultant large lepton asymmetry can affect the primordial helium abundance.
The experimental and cosmological constraints on the  U(1)$_{L_\mu-L_\tau}$ gauge boson mass and gauge coupling compiled in Ref.~\cite{Bernal:2025szh}  
motivates $m_X=\mathcal{O}(10\text{--}100)~{\rm MeV}$.  
For our benchmark coupling, $g_{L_\mu-L_\tau}=3\times10^{-4}$, we take $m_X\simeq10~{\rm MeV}$. 
In the previous study by some of the authors~\cite{Hu:2025nep}, we focused on the generation of a large lepton asymmetry that could account for the primordial helium abundance suggested recent by CMB observations such as ACT~\cite{AtacamaCosmologyTelescope:2025blo} and observations of the extremely metal-poor galaxies~\cite{Yanagisawa:2025mgx}. 
They found that $\eta_H \simeq 10^{-3}$ with $H_\mathrm{inf}<10^{-1}$ GeV 
may explain these observations. 
However, the baryon asymmetry generated before the electroweak symmetry breaking through the dissipation of magnetic helicity was not fully taken into account in that study. 
We find that the baryon and lepton asymmetries generated in this scenario have opposite signs. 
In particular, 
the positive lepton asymmetry required to explain the primordial helium abundance suggested by these observations is accompanied by a negative baryon asymmetry. 
Requiring this negative contribution to be smaller than the observed baryon asymmetry, $|\eta_B| < \eta_B^\mathrm{obs}$, imposes a stronger constraint,  $H_\mathrm{inf}<10^{-7}$ GeV, as shown in Fig.~\ref{parameterspace}. 
This constraint could be relaxed if 
another mechanism generates a compensating positive baryon asymmetry or if the helicity dissipation before the electroweak symmetry breaking is further suppressed. 
Note that reversing the sign of $\eta_H$ reverses the sign of both the resultant baryon and lepton numbers.

There are several issues that deserve further investigation.  
In this study, we have assumed that 
the Schwinger effect during inflation is strongly suppressed by large induced masses.
If the charged fermions remain light, Schwinger pair production induces a current that screens the amplified electric field and changes the gauge-field spectrum~\cite{Domcke:2018eki,Gorbar:2021rlt,vonEckardstein:2024tix}.  
Moreover, since the anomaly equation also applies  during inflation, a lepton asymmetry associated with the $L_\mu - L_\tau$ magnetic helicity is already  generated by the end of inflation. 
For our benchmark coupling, 
Ref.~\cite{Hu:2025nep} determined an upper bound of the magnetic helicity,
$\eta_H\lesssim1.4\times10^{-6}$,  
by requiring that the energy density of gauge fields and fermions remain sufficiently subdominant to the inflaton energy density.
Although the large lepton asymmetry suggested by recent helium observations cannot be obtained in this case, the observed baryon asymmetry may still be reproduced in some regions of parameter space. 
Determination of these regions, however, requires a self-consistent treatment of the coupled evolution of the particle asymmetries and magnetic fields, including the chiral MHD effects~\cite{Brandenburg:2023rul,Brandenburg:2023aco}.

Another uncertainty concerns the detailed dynamics of the decay of magnetic helicity and lepton number generation at U(1)$_{L_\mu-L_\tau}$ breaking.  
Following Ref.~\cite{Fukuda:2024pkh}, 
we may suppose that the magnetic fields rapidly dissipates  or decays due to the emergence of the gauge boson mass, generating a lepton asymmetry.
However, the actual dynamics may be more complicated. 
For example, local cosmic strings may form at the symmetry breaking, with the magnetic field flux confined into them, keeping the helical structure. 
The subsequent evolution and reconnection of the cosmic string network could lead to helicity decay and lepton asymmetry generation, although the relevant timescale may differ from that assumed in the analysis of Ref.~\cite{Fukuda:2024pkh}. 
Moreover, depending on the 
dynamics during the U(1)$_{L_\mu-L_\tau}$ higgsing, the strings may decay 
accompanying the change of the higgs winding number, {\it e.g.,} generating monopole-antimonopole pairs~\cite{Hamada:2025cwu,Fukuda:2025nmc}, 
in which case the lepton asymmetry is not necessarily generated, 
since the notion of magnetic helicity (and its decay) becomes obscure when the higgs winding changes.
A more detailed investigation of these processes is needed to determine the efficiency and timescale of lepton asymmetry generation and is left for future study.

%%%%%%%%%%%%%%%%%%%%%%%%%%%%%%%%%%%%%%%%%%%%
\acknowledgments
%%%%%%%%%%%%%%%%%%%%%%%%%%%%%%%%%%%%%%%%%%%%

The work of K.K.~was supported by the National Natural Science Foundation of China (NSFC) under Grant No.~W2532007 and 12547104, and by JSPS KAKENHI Grant-in-Aid for Challenging Research (Exploratory) JP23K17687. 
The work of F.U.~was supported by IBS under the project code
IBS-R018-D3.

\bibliographystyle{JHEP}
\bibliography{papers}

\end{document}